\documentstyle[12pt,epsf]{article}

\newcommand{\bmat}{\left(\begin{array}}
\newcommand{\emat}{\end{array}\right)}

\def\yzero{\smash{\hbox{$y\kern-4pt\raise1pt\hbox{${}^\circ$}$}}}

\def\-{\hphantom{-}}

\def\s2{\frac{1}{\sqrt2}}

\def\beq{\begin{equation}}
\def\eeq{\end{equation}}
\def\beqa{\begin{eqnarray}}
\def\eeqa{\end{eqnarray}}
\def\tr{{\rm tr \,}}
\def\Tr{{\rm Tr \,}}
\def\diag{{\rm diag \,}}
\def\IF{\relax{\rm I\kern-.18em F}}
\def\II{\relax{\rm I\kern-.18em I}}
\def\IP{\relax{\rm I\kern-.18em P}}
\def\inbar{\vrule height1.5ex width.4pt depth0pt}
\def\IC{\relax\hbox{\kern.25em$\inbar\kern-.3em{\rm C}$}}
\def\IR{\relax{\rm I\kern-.18em R}}

\def\cc{{\cal C}}
\def\ck{{\cal K}}

\def\cm{{\cal M}}

\def\NN{{\cal N}}
\def\Dsl{\,\raise.15ex\hbox{/}\mkern-13.5mu D} 
\def\IZ{Z}

\def\--{O$6^{-}$-O$6^{-}$}
\def\+-{O$6^{+}$-O$6^{-}$}

\newcommand{\drawsquare}[2]{\hbox{%
\rule{#2pt}{#1pt}\hskip-#2pt
\rule{#1pt}{#2pt}\hskip-#1pt
\rule[#1pt]{#1pt}{#2pt}}\rule[#1pt]{#2pt}{#2pt}\hskip-#2pt
\rule{#2pt}{#1pt}}

\newcommand{\fund}{\raisebox{-.5pt}{\drawsquare{6.5}{0.4}}}
\newcommand{\Ysymm}{\raisebox{-.5pt}{\drawsquare{6.5}{0.4}}\hskip-0.4pt%
        \raisebox{-.5pt}{\drawsquare{6.5}{0.4}}}
\newcommand{\Yasymm}{\raisebox{-3.5pt}{\drawsquare{6.5}{0.4}}\hskip-6.9pt%
        \raisebox{3pt}{\drawsquare{6.5}{0.4}}}

\topmargin
-1.5cm
\textwidth
15.5cm
\textheight
24cm
\oddsidemargin
0.7cm
\evensidemargin
1.2cm

\begin{document}

\makeatletter
\@addtoreset{equation}{section}
\makeatother
\renewcommand{\theequation}{\thesection.\arabic{equation}}
\pagestyle{empty}
\rightline{IASSNS-HEP-98/76}
\rightline{\tt hep-th/9808161}
\vspace{0.5cm}
\begin{center}
\LARGE{A Note on Superconformal $\NN=2$ theories and Orientifolds
\\[10mm]}
\large{Jaemo ~Park \footnote{E-mail: jaemo@sns.ias.edu}, \hspace{.3cm} 
Angel~M.~Uranga \footnote{E-mail: uranga@sns.ias.edu} \\[2mm]}
{\em School of Natural Sciences, Institute for Advanced Study \\
Princeton NJ 08540,USA.\\[1mm]}

\vspace*{2cm}

\small{\bf Abstract} \\[7mm]
\end{center}

\begin{center}
\begin{minipage}[h]{14.0cm}

{\small We construct the T duals of certain type IIA brane configurations 
with one compact dimension (elliptic models) which contain orientifold 
planes. These configurations realize four-dimensional $\NN=2$ finite field 
theories. For elliptic models with two negatively charged orientifold 
six-planes, the T duals are given by D3 branes at singularities in the 
presence of O7-planes and D7-branes. For elliptic models with two 
oppositely charged orientifold planes, the T duals are D3 branes at a 
different kind of orientifold singularities, which do not require D7 
branes. We construct the adequate orientifold groups, and show that the 
cancellation of twisted tadpoles is equivalent to the finiteness of the 
corresponding field theory. One family of models contains orthogonal 
and symplectic gauge factors at the same time. These new orientifolds can 
also be used to 
define some six-dimensional RG fixed points which have been discussed from 
the type IIA brane configuration perspective.}
\end{minipage}
\end{center}
\newpage
\setcounter{page}{1}
\pagestyle{plain}
\renewcommand{\thefootnote}{\arabic{footnote}}
\setcounter{footnote}{0}

\section{Introduction}

There are basically two kinds of brane configurations in string theory
that have been used to study four-dimensional $\NN=2$ gauge theories. The
first, introduced in \cite{wit4d} in the spirit of \cite{hw}, makes
use of
sets of type IIA D4 branes suspended between NS branes. Also, D6 branes
may be added without breaking further supersymmetries. The gauge theory is
realized in the non-compact part of the D4 brane world-volume. This
picture is very intuitive geometrically, and has been particularly useful
in finding the exact solution of a large class of $\NN=2$ theories. Also,
it provides a simple construction of superconformal theories. A
particularly nice such family, the so-called elliptic models, can
be obtained upon taking the coordinate $x^6$, along which the D4 branes
are finite, to be compact.

A second type of construction is realized by type IIB D3 branes probing a
certain background \cite{bds,dl}. One simple example is that of a
set of D3 branes sitting at an $A_k$ singularity \cite{dm}. In fact,
this system is
related to the elliptic models described above by a T-duality along the
compact direction in the type IIA brane configuration \cite{lust}.
Even though this picture has not proved so suitable for solving the
$\NN=2$ theories, it has several advantages. First, since it does not
contain any NS branes, it can be analyzed perturbatively using string
theory techniques. Also, it allows a simple realization of superconformal
theories, by considering backgrounds in which the type IIB coupling
constant (which controls the gauge theory coupling constant) does not
depend on the position in spacetime. Finally, the near horizon geometry
in these configurations is a quotient of $AdS_5\times S^5$, and allows the
use of the AdS/CFT correspondence \cite{ads} to study their  
supergravity/string theory description of their large $N$ limit.

Other families of elliptic models can be obtained by the introduction of
orientifold six-planes \cite{uranga} (configurations with O4-planes 
\cite{lll,bsty} will not be studied in the present paper). It is a 
natural question what their T-dual type IIB versions are. 
There are two main classes of
models, depending on the charges of the O6-planes. If both are
negatively charged, (\-- configuration) finiteness of the
$\NN=2$ field theory requires the
presence of D6 branes. Their T-duals are certain type IIB orientifolds,
containing O7-planes and D7 branes (a particular case in this
family reproduces the F-theory background introduced in \cite{sen},
which was used in \cite{bds,aharony,dls} to the study of some $\NN=2$ 
theories). These are analogous to the orientifolds introduced in 
\cite{dm,intri,bi1} to study six-dimensional theories.

The remaining case has oppositely charged O6-planes (\+- configuration) 
\footnote{The configuration with two positively charged orientifolds 
cannot lead to finite theories.}. Finiteness of the
field theory on the D3 branes requires in this case the absence of D6
branes. The T-dual of these configurations has not been determined, and
the purpose of this note is to construct such type IIB orientifolds. They
will have the property that cancellation of tadpoles is achieved without
D7 branes.

In all cases we will show that cancellation of twisted tadpoles is 
equivalent to the finiteness of the field theory. This follows from the 
fact that tadpoles are sources for twisted closed string modes, which 
propagate in two dimensions. The logarithmic solutions to the 
corresponding Laplace equation correspond to the one-loop evolution of the 
gauge coupling with the scale. Thus, finiteness of the $\NN=2$ theory 
follows from tadpole cancellation.

We would like to stress that brane configurations T-dual to those yielding
the finite four dimensional theories have been used to construct 
consistent
six-dimensional field theories in \cite{bk1,bk2,hz} \footnote{In the 
six-dimensional case, twisted modes have no dimensions to propagate. 
Thus, tadpole cancellation is a {\em consistency} requirement, rather than 
a choice.}. They contain 
NS fivebranes, D6 and D8 branes, and possibly O6- and O8-planes. By
T-dualizing along the compact dimension, the configurations with two
negatively charged O8-planes are related to D5 branes at
the orientifold singularities introduced in \cite{dm,intri,bi1}. The
orientifolds we will construct in Section~3 can be used to provide the
T-dual of the brane configurations with oppositely charged O8-planes.
This gives an alternative construction for the six-dimensional field
theories introduced in \cite{bk2,hz}. Comments concerning the
six-dimensional version of these models will be made throughout the paper.

The structure of this paper is as follows. In Section~2 we review the
brane constructions of the elliptic models with orientifold planes. In
Section~3 we briefly comment how the orientifolds in \cite{bi1}
provide the T-dual version of the models with two negatively charged
orientifold six-planes. In Section~4 we describe the type IIB orientifolds
which are T-dual to the \+- configuration, and show explicitly that 
cancellation of twisted tadpole implies the finiteness of the 
four-dimensional $\NN=2$ theories. Section~5 contains our conclusions.

\section{Description of the brane configurations}

We will be studying brane configurations of NS branes along 012345, D4
branes along 01236, and D6 branes along 0123789 \cite{wit4d}. The D4 
branes will have
finite extent along the direction $x^6$, which is taken to be compact.
Orientifold six- and four-planes, if present, are parallel to the D branes
of the corresponding dimension. The brane models with O6- and
O4-planes have been discussed in \cite{uranga} and \cite{lll,bsty},
respectively.

Let us center the models with O6-planes. Since the 
direction $x^6$ is compact, there are two orientifold planes. The 
configuration in the
cover space is a set of $N$ NS branes located at points in the circle
parametrized by $x^6$. In the $i^{th}$ interval they define, there are
$v_i$ D4 branes suspended between NS branes, and $w_i$ D6 branes. The
whole brane configuration must be invariant under the $\IZ_2$ symmetry
inverting the coordinates 456. The basic rules to read the spectrum are
discussed in \cite{uranga,bk2,hz}, using previous results in
\cite{landlop}.

\begin{figure}
\centering
\epsfxsize=4.5in
\hspace*{0in}\vspace*{.2in}
\epsffile{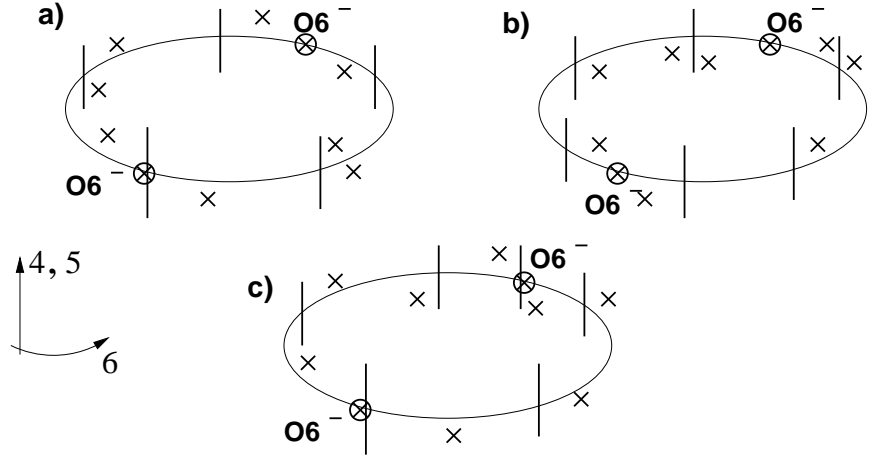}
\caption{\small The three families of brane configurations in
the background of two negatively charged O6-planes. The short vertical
lines represent the NS branes, the crossed circles are the orientifold
planes, while uncircled crosses denote the D6 branes. The D4 branes 
stretch in the interval along the circle in $x^6$. In order to yield 
finite theories, the number of D4 branes is generically different at each 
interval. For the sake of clarity we have not attempted to draw them.}
\label{fig:minusminus}
\end{figure}

When both O6-planes are negatively
charged, finiteness of the field theory requires the presence of 8 D6
branes, $\sum_{j} w_j=8$. Notice that this condition is equivalent to the 
cancellation of the RR charge in the brane picture. There are three
main families within this class, depending on the positions of the NS
branes.

{\bf i)} The number of NS branes is odd, $N=2P+1$. One typical brane
configuration is depicted in Figure~1a. Notice that the $\IZ_2$ symmetry
forces one of the NS branes to intersect one O$6^-$. It also requires 
$v_{-i}=v_i$, $w_{-i}=w_i$. The gauge theory 
\footnote{Our convention is that the fundamental representation of 
$USp(k)$ has dimension $k$.} is 
\beqa
& USp(v_0)\times SU(v_1) \times \ldots \times SU(v_P) & \nonumber\\
& \bigoplus_{j=0}^{P-1} (\fund_j,\fund_{j+1}) + \Yasymm_P + \frac 12
w_0\fund_0 + \bigoplus_{j=1}^{P} w_j \fund_j &
\label{spec1}
\eeqa
where the subindices denote the corresponding group factor. In this and 
the following spectra we have taken into account that the $U(1)$ factors 
are frozen at low energies \cite{wit4d}.

The condition for this theory to be finite is the vanishing of the
one-loop beta functions,
\beq
b_j \; \equiv\;
-2v_j+v_{j-1}+v_{j+1}+w_j-4\delta_{j,0}-2\delta_{j,P}-2\delta_{j,P+1}\,
=\, 0.
\label{beta1}
\eeq
From the brane configuration point of view, these conditions are obtained 
by requiring the linking numbers of all the NS branes to be equal (so 
that their asymptotic bending is identical). A similar comment applies to 
all other models. 

{\bf ii)} The number of NS branes is even, $N=2P$, and there are no NS
branes intersecting the O6-planes. One example is shown in Figure~1b.
Here we again have $v_{-i}=v_i$, $w_{-i}=w_i$.
The gauge theory in this family of models is
\beqa
& USp(v_0)\times SU(v_1) \times \ldots \times SU(v_{P-1})\times USp(v_P) &
\nonumber\\
& \bigoplus_{j=0}^{P-1} (\fund_j,\fund_{j+1}) + \frac 12 w_0\fund_0 +
\bigoplus_{j=1}^{P-1} w_j \fund_j + \frac 12 w_P\fund_P &
\label{spec2}
\eeqa
The finiteness conditions read
\beq
b_j \; \equiv\; -2v_j+v_{j-1}+v_{j+1}+w_j-4\delta_{j,0}-4\delta_{j,P} \, 
=\, 0.
\label{beta2}
\eeq

{\bf iii)} The number of NS branes is even, $N=2P$, but two NS branes
intersect the O6-planes (see Figure~1c). The $Z_2$ symmetry imposes 
$v_{-i+1}=v_i$, $w_{-i+1}=w_i$. The gauge theory is
\beqa
& SU(v_1)\times SU(v_2) \times \ldots \times SU(v_{P-1})\times SU(v_P) &
\nonumber\\
& \Yasymm_{\, 1} +\bigoplus_{j=1}^{P-1} (\fund_j,\fund_{j+1}) + \Yasymm_P
+ \bigoplus_{j=1}^{P} w_j \fund_j &
\label{spec3}
\eeqa
In this case, finiteness is achieved when
\beq
b_j \; \equiv\; -2v_j+v_{j-1}+v_{j+1}+w_j -2\delta_{j,1} -2\delta_{j,2P}
-2\delta_{j,P} -2\delta_{j,P+1} \, =\, 0.
\label{beta3}
\eeq

\begin{figure}
\centering
\epsfxsize=4.5in
\hspace*{0in}\vspace*{.2in}
\epsffile{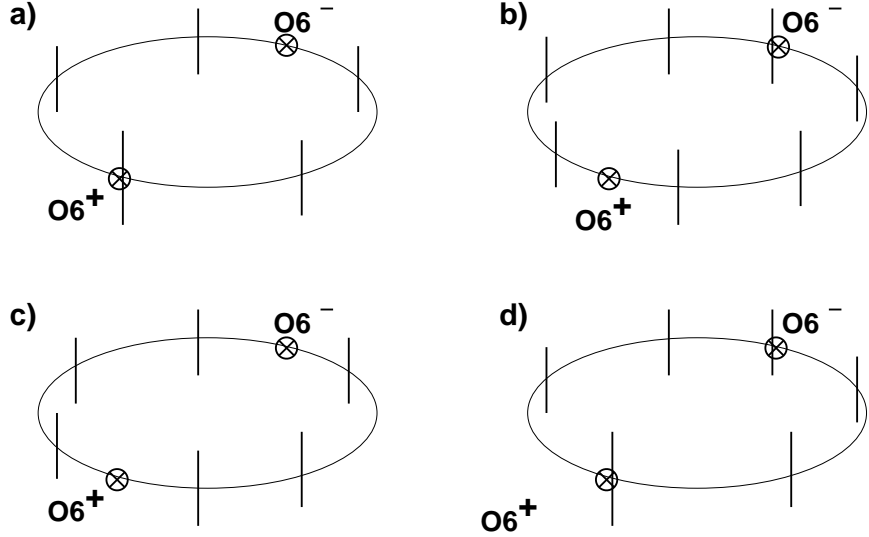}
\caption{\small The four families of theories arising in the
\+- configuration.}
\label{fig:plusminus}
\end{figure}

\medskip

For the \+- configuration the RR charge already cancels, so finiteness 
does not allow the presence of D6
branes, thus $w_j=0$. There are four families of models depending on the
positions of the NS branes.

{\bf i)} The number of NS branes is odd, $N=2P+1$, and the unpaired NS
brane intersects the O$6^+$-plane. One such example is shown in Figure~2a. 
The resulting gauge theory is
\beqa
& USp(v_0)\times SU(v_1) \times \ldots \times SU(v_P) & \nonumber\\
& \bigoplus_{j=0}^{P-1} (\fund_j,\fund_{j+1}) + \Ysymm_P &
\label{spec1p}
\eeqa
The finiteness conditions are
\beq
b_j \; \equiv\;
-2v_j+v_{j-1}+v_{j+1} -4\delta_{j,0}+2\delta_{j,P}+2\delta_{j,P+1}\, =
\, 0.
\label{beta1p}
\eeq
Comparing them with (\ref{beta1}) we clearly see the effect of the sign of
the orientifold six-planes in the signs of the last contributions. A
similar comment applies to the remaining models.

These conditions constrain the gauge group to be $USp(2k)$$\times 
SU(2k+2)$$\times \ldots $$\times$$ SU(2k+2P)$.

{\bf i')} The number of NS branes is odd, $N=2P+1$, and the unpaired NS
brane intersects the O$6^-$-plane, as depicted in Figure~2b.
The field theory spectrum is
\beqa
& SO(v_0)\times SU(v_1) \times \ldots \times SU(v_P) & \nonumber\\
& \bigoplus_{j=0}^{P-1} (\fund_j,\fund_{j+1}) + \Yasymm_P &
\label{spec1pp}
\eeqa
The vanishing of the beta functions reads
\beq
b_j \; \equiv\; -2v_j+v_{j-1}+v_{j+1} +4\delta_{j,0} -2\delta_{j,P}
-2\delta_{j,P+1}\,=\, 0.
\label{beta1pp}
\eeq
The most general solution has gauge group $SO(n)$$\times SU(n-2)$$\times 
\ldots$$\times SU(n-2P)$.

{\bf ii)} The number of NS branes is even, $N=2P$, and there are no NS
branes intersecting the O6-planes (see Figure~2c).
The gauge theory is
\beqa
& SO(v_0)\times SU(v_1) \times \ldots \times SU(v_{P-1})\times USp(v_P) &
\nonumber\\
& \bigoplus_{j=0}^{P-1} (\fund_j,\fund_{j+1}) &
\label{spec2p}
\eeqa
The finiteness conditions in this case are
\beq
b_j \; \equiv\; -2v_j  +v_{j-1} +v_{j+1} +4\delta_{j,0} -4\delta_{j,P}\, 
=\, 0.
\label{beta2p}
\eeq
The finite theory has gauge group $SO(2k)$$\times SU(2k-2)$$\times 
\ldots$$\times SU(2k-2P+2)$$\times$$USp(2k-2P)$.

{\bf iii)} The number of NS branes is even, $N=2P$, but two NS branes
intersect the O6-planes, as in Figure~2d.
This construction yields the spectrum
\beqa
& SU(v_1)\times SU(v_2) \times \ldots \times SU(v_{P-1})\times SU(v_P) &
\nonumber\\
& \Yasymm_{\, 1} +\bigoplus_{j=1}^{P-1} (\fund_j,\fund_{j+1}) + \Ysymm_P &
\label{spec3p}
\eeqa
Finite field theories are obtained when
\beq
b_j \; \equiv\; -2v_j+v_{j-1}+v_{j+1} -2\delta_{j,1} -2\delta_{j,2P}
+2\delta_{j,P} +2\delta_{j,P+1}\, = \, 0.
\label{beta3p}
\eeq
These conditions imply the group is $SU(n)$$\times\ldots$$\times 
SU(n+2P-2)$.

\medskip

These field theories can also be considered in six dimensions
\footnote{Being more precise, one should also include the dynamics of the
NS branes in this case \cite{bk2,hz}.}.
The conditions for cancellation of six-dimensional
anomalies are obtained from the four-dimensional finiteness conditions by
merely replacing $2 \to 8$, and $4 \to 16$ in the Kronecker delta
contributions of the quantities $b_j$. Also, in the O$8^-$-O$8^-$ 
configuration, we must require the presence of 32 D8 branes, $\sum_{j} 
w_j=32$.

\section{The \-- configuration}

In this section we discuss the type IIB orientifolds T dual to the brane 
configurations with both O6-planes negatively charged. 
If we T-dualize along the $x^6$-direction, the two O6-planes are mapped
to one O7-plane located at $x^4=x^5=0$, say.
Also the $N$ NS5-branes are mapped to $Z_N$-singularity located at
$x^6=x^7=x^8=x^9=0$. The D4-branes are turned into D3-branes.
So the natural guess is that the T-dual of the O$6^{-}$-O$6^{-}$ 
configuration is just the composition of the individual T-duality maps.
The T-dual is thus described by the D3-branes with the orientifold
projection,
\begin{equation}
(1+\theta+\theta^2+\cdots +\theta^{N-1})(1+\Omega').
\end{equation}
where $\theta$ is the generator of 
$Z_{N}$ action
\begin{eqnarray}
z_1& \rightarrow & e^{\frac{2\pi i}{N}}z_1  \\
z_2& \rightarrow & e^{-\frac{2\pi i}{N}}z_2
\end{eqnarray}
with $z_1=x^6+ix^7$ and $z_2=x^8+ix^9$. Also, $\Omega'=\Omega R_{45} 
(-1)^{F_L}$, where $R_{45}$ is a reflection in 
$x^4,x^5$-directions and $(-1)^{F_L}$ acts as $-1$ on the Ramond sector 
of the left movers.  
For one O7-plane,we need 8 D7-branes in order to neutralize the
RR-charge of the O7-plane.

If we compactify $x^4, x^5$-directions and T-dualize along
these directions, we have a system of D5-branes with the orientifold 
projection
\begin{equation}
(1+\theta+\theta^2+\cdots +\theta^{N-1})(1+\Omega ).
\end{equation}
Blum and Intriligator have already considered this orientifold projection
for the realization of six-dimensional field theories,
and the three cases of O$6^{-}$-O$6^{-}$ have counterparts in their
construction. So our discussion will be brief and mainly refer to the
construction in \cite{bi1} \footnote{Compact six-dimensional examples 
involving orientifold projections of this kind have been considered in 
\cite{gp,gj,dp}. Also, some particular examples of D3 branes at these 
orientifold singularities have appeared in \cite{kakush1}.}.

\subsection{The odd order case}

The spectrum (\ref{spec1}) can be reproduced by the following choice of 
Chan-Paton matrices. We have
\begin{equation}
\gamma_{\theta ,3}=\diag (1_{v_0},\theta 1_{v_1}, \cdots \theta^{P}
1_{v_P},\theta^{P+1} 1_{v_{P}}, \cdots \theta^{2P}1_{v_1})
\end{equation}
(with $\theta=e^{\frac{2\pi i}{N}}$) and a similar expression for 
$\gamma_{\theta,7}$. Here we choose $\gamma_{\theta^k}=(\gamma_{\theta})^k$.
Also we have
\begin{equation}
\gamma_{\Omega',3}={\small \left( \begin{array}{ccccccccc}
\varepsilon_{v_0} & & & & & & & & \\
 & & & & & & & & 1_{v_1} \\
 & & & & & & &1_{v_2} &  \\
 & & & & & & \cdots& &  \\
 & & & & & 1_{v_P} & & & \\
 & & & &-1_{v_P}& & & & \\
 & & & \cdots & & & & & \\
 & & -1_{v_2}& & & & & & \\
 & -1_{v_1} & & & & & & &  \end{array}  \right)}
\end{equation}
where 
\begin{equation}
\varepsilon_{v_0}={\small \left( \begin{array}{cc}
0 & 1_{\frac{v_0}{2}} \\
-1_{\frac{v_0}{2}} & 0   \end{array}  \right)}.
\end{equation}
For D7 branes, $\gamma_{\Omega',7}$ has the same block structure, but is 
symmetric ({\em i.e.} $1_{v_0}$ replaces $\varepsilon_{v_0}$, and all 
unit matrices appear with positive sign). These matrices yield 
the spectrum shown in (\ref{spec1})

For this case the twisted tadpole is \cite{gj,dp}
\begin{equation}
\sum_{k=1}^{N-1} \frac{1}{4\sin^2\frac{2\pi k}{N}}(\sum_{j} 
w_{j}e^{\frac{4\pi i kj}{N}}-4\sin^2\frac{2\pi k}{N}\sum_{j}  
v_{j}e^{\frac{4\pi i kj}{N}} -8\cos^2\frac{\pi k}{N})^2=0.
\label{eq:tad1}
\end{equation}
Here $\tr\gamma_{\theta^{2k},7}=\sum_j w_{j}e^{\frac{4i\pi kj}{N}}$
and $\tr\gamma_{\theta^{2k,3}}=\sum_j v_{j}e^{\frac{4i\pi kj}{N}}$.

By expanding the sine and cosine functions in exponentials, it can be 
shown that the 
equations of cancellation of tadpoles are equivalent to the finiteness 
conditions (\ref{beta1}).

The difference between (\ref{eq:tad1}) and the tadpole expression 
in the six-dimensional context of \cite{bi1} is that
we have 8 instead of 32 in the last term of the tadpole equation.
This factor-4 difference reflects the T-dual relation between our model
and six-dimensional model considered in \cite{bi1}. 
The crosscap states are products of the crosscap coming from the
twisted action and the crosscap coming from the toroidal direction.
The part of the crosscap from the twisted part remains the same
and the crosscap from the toroidal direction shows the usual
behavior under the T-duality. That is the origin of the factor-4
difference. Same thing happens for the corresponding orientifolds in cases 
ii) and iii) to be discussed below.

\subsection{The even order cases}

One apparently curious fact is that the brane configurations ii) and iii)
lead to the same orientifold projection. The difference has been discussed 
in the six-dimensional context \cite{bk2,hz}, as we presently discuss 
\footnote{The argument follows for the four-dimensional case by replacing 
`tensor multiplet' by `vector multiplet'.}. 

Gauge couplings in six-dimensional $N=1$ 
theories belong to tensor multiplets. In the brane configuration, the 
gauge coupling is proportional to the distance between two NS5-branes in 
$x^6$-direction. On the other hand, motion in $x^7,x^8,x^9$-directions 
of a NS-brane correspond to a hypermultiplet. 

In the configuration ii), $P$ NS5-branes are located within the 
interval  between two O6-planes. By choosing one particular 
NS-brane, we can take the independent 
gauge couplings to be the distance of the remaining $P-1$ NS-branes  
and the particular NS-brane and the distance between this particular 
NS-brane and its mirror image. 
Hence, we have $P$ tensor multiplets. We can also move NS5-branes in 
$x^7,x^8,x^9$-directions pairwise under the orientifolding. 
Excluding the overall motion of the $P$ pairs, we have P$-$1 
hypermultiplets. 

In the configuration iii), we have $P-1$ NS-branes located within the 
interval between two O6-planes. In addition, we have a  
NS-brane stuck at each O6-plane. We can take the independent gauge 
couplings to be the distance of the $P-1$ NS-branes and one of the stuck 
branes. The distance between the two stuck branes is fixed and cannot be 
an independent gauge coupling. Thus we get $P-1$ tensor multiplets.
On the other hand, motion of the $P-1$ pairs of NS branes along 
$x^7,x^8,x^9$ contribute $P-2$ hypermultiplets. The independent motions 
of the unpaired branes contributes two more hypermultiplets. We have P 
hypermultiplets in total. 

Thus the configuration ii) lead to $P$ tensor multiplets and $P-1$ 
hypermultiplets, while configuration iii) leads to $P-1$ 
tensor multiplets and $P$ hypermultiplets.

These two possibilities are indeed discussed in \cite{bi1} in the T-dual 
type IIB orientifold 
version. The difference between both models is that for $Z_2$ twisted 
sector, we can keep either the hypermultiplet or the tensor multiplet 
under the orientifold projection. According to \cite{bi1},
this fact is related to the distinction between Type I configurations with 
(possible) vector structure and without vector structure.
This distinction originates from the fact that the gauge group
of Type I string theory is $Spin(32)/Z_2$. Thus if we
choose the Chan-Paton matrices $\gamma$ for the $Z_N$ representation,
$\gamma^N=1$ in $Spin(32)/Z_2$. Thus we can have $\gamma^N=1$ or
$\gamma^N=w$ in $Spin(32)$ where $w$ is the generator of $Z_2$
in $Spin(32)/Z_2$. The different action on the $Z_2$ twisted sector
leads to different Chan-Paton matrices corresponding to
a configuration with (without) vector structure. 

The number of tensor multiplets and hypermultiplets
of the configuration  ii) matches the spectrum with vector structure and 
the configuration iii) matches the spectrum without vector structure. 
The gauge group and matter content on the D-brane world-volume 
also agree with this identification, as we will see shortly.
Finally, we anticipate that the argument above also applies to 
O$6^{+}$-O$6^{-}$ configurations.

\medskip
 
For the case with vector structure, the Chan-Paton matrices have the 
structure
\begin{equation}
\gamma_{\theta,3}=\diag(1_{v_0},\theta 1_{v_1}, \cdots,\theta^{P-1} 
1_{v_{P-1}} \theta^{P} 1_{v_P}, \theta^{P+1} 1_{v_{P-1}}, \cdots 
\theta^{2P-1}1_{v_1})
\end{equation}
(and analogously for $\gamma_{\theta,7}$) and
\begin{equation}
\gamma_{\Omega',3}={\small \left( \begin{array}{cccccccc}
\varepsilon_{v_0} & & & & & & &  \\
  & & & & & & & 1_{v_1} \\
  & & & & & &\cdots &  \\
  & & & & & 1_{v_{P-1}}& &  \\
  & & & & \varepsilon_{v_P} & & & \\
  & & &-1_{v_{P-1}}& & & & \\
  & & \cdots & & & & & \\
  & -1_{v_1} & & & & & &  \end{array}  \right)}. 
\end{equation}
(and a symmetric version of this for $\gamma_{\Omega',7}$).
These matrices reproduce the spectrum (\ref{spec2})

Notice that these matrices have the property
\beqa
\Tr (\gamma_{\theta^{k}\Omega'}^{T} \gamma_{\theta^{k}\Omega'}^{-1})
\, = \, \mp \Tr (\gamma_{(\theta)^{2k}}) \, =\, \Tr
(\gamma_{\theta^{k+P} \Omega'}^{T} \gamma_{\theta^{k+P}\Omega'}^{-1})
\label{prop2}
\eeqa
with the upper (lower) sign for D3 (D7) branes. The positive relative sign 
between the first and third terms characterizes the action of the $\IZ_2$
twist to correspond to models with vector structure \cite{bi1}. This is 
also reflected in a positive relative sign between the 
untwisted and $Z_2$-twisted contributions to the Klein bottle. 

The twisted tadpole equation is
\begin{equation}
\sum_{k=1}^{N-1} \frac{1}{4\sin^2\frac{\pi k}{N}} \left[\sum_{j} w_{j}
e^{\frac{2i\pi  kj}{N}} -4\sin^2\frac{\pi k}{N} \sum_j v_{j} 
e^{\frac{2i\pi  kj}{N}} -8\delta_{k,0\, {\rm mod}\, 2} \right]^2=0.
\label{eq:tad3}
\end{equation}

It is easy to show that the cancellation of tadpoles is equivalent to the 
finiteness conditions (\ref{beta2}).

\medskip

For $N=2P$ without vector structure, the Chan-Paton matrices are given by 
\begin{equation}
\gamma_{\theta,3}=e^{-\frac{\pi i}{N}}\diag(e^{\frac{2\pi i}{N}}1_{v_1},
e^{\frac{4\pi i}{N}}1_{v_2}, \cdots e^{\frac{2\pi i P}{N}}1_{v_P},
e^{\frac{2\pi i(P+1)}{N}}1_{v_P}, \cdots,e^{\frac{2\pi i(2P-1)}{N}}1_{v_2}
e^{\frac{2\pi i(2P)}{N}}1_{v_1})
\end{equation}
(analogously for $\gamma_{\theta,7}$) and
\begin{equation}
\gamma_{\Omega',3}={\small \left( \begin{array}{cccccccc}
  & & & & & & & 1_{v_1} \\
  & & & & & &1_{v_2} &  \\
  & & & & & \cdots& &  \\
  & & & & 1_{v_P} & & & \\
  & & &-1_{v_P}& & & & \\
  & & \cdots & & & & & \\
  & -1_{v_2}& & & & & & \\
  -1_{v_1} & & & & & & &  \end{array}  \right)}
\end{equation}
(and a symmetric version for $\gamma_{\Omega',7}$)

These matrices yield the field theory spectrum in (\ref{spec3}).
The matrices satisfy
\beqa
\Tr (\gamma_{\theta^{k}\Omega'}^{T} \gamma_{\theta^{k}\Omega'}^{-1})
\, = \, \mp \Tr (\gamma_{(\theta)^{2k}}) \, =\, -\Tr
(\gamma_{\theta^{k+P} \Omega'}^{T} \gamma_{\theta^{k+P}\Omega'}^{-1})
\label{prop3}
\eeqa
with the upper (lower) sign for D3 (D7) branes. The relative minus sign
between the first and third contributions implies these are models without 
vector structure. 

The tadpole equation can be taken directly from \cite{gj,dp}, and read
\begin{equation}
\sum_{k=1}^{N-1} \frac{1}{4\sin^2\frac{\pi k}{N}}\left[\sum_{j} w_{j}
e^{\frac{\pi i(2j-1)k}{N}} -4\sin^2\frac{\pi k}{N} \sum_{j} v_{j}
e^{\frac{\pi i(2j-1)k}{N}}-8\delta_{k,0\,{\rm mod}\, 2}
\cos\frac{\pi k}{N} \right]^2=0.
\label{eq:tad2}
\end{equation}
We have different choice of $\gamma_{\theta}$ and different tadpole 
expression from those used in \cite{bi1}. Basically we absorb the 
phase factor appearing in their tadpole expression into the redefinition 
of the $\gamma$ matrices. 

Again, the tadpole cancellation conditions can be recast as the vanishing 
of the beta functions for the corresponding four-dimensional $\NN=2$ field 
theory, eq (\ref{beta3}).

\medskip

Note that all the above Chan-Paton matrices satisfy the constraint
imposed by Polchinski \cite{polchinski}. That is, if $R$ denotes the 
$\IZ_2$ twist, we have
\begin{equation}
\gamma_R=\mp\gamma_{\Omega'}\gamma_R^T \gamma_{\Omega'}^{-1} 
\label{eq:p1}
\end{equation}
with the negative (positive) sign for models that keep the hypermultiplet 
(tensor multiplet) in their $Z_2$ twisted sector.

Conversely if we know $\gamma_R$ and $\gamma_{\Omega}$, 
all Chan-Paton matrices can be determined from the tadpole
equation and the Chan-Paton algebra relation.
Thus the conditions (\ref{prop2}), (\ref{prop3})
are the corollary of (\ref{eq:p1}).

It is interesting to note that in six dimensions, the above orientifold 
construction gives theories free of six-dimensional anomaly, while the 
T-dual construction in four dimensions gives finite N=2 gauge theories. 
This nicely illustrates how tadpole conditions encode the relevant quantum 
effects of the field theory.

\section{The O$6^{+}$-O$6^-$ configuration}

The main feature of the brane configurations with oppositely charged
orientifold planes is the absence of D6 branes. This fact follows from
finiteness in the field theory, or from cancellation of RR charge. The
natural interpretation in terms of the T-dual picture of D3 branes at
orientifold singularities is that tadpole cancellation
in the Type IIB orientifolds is achieved without D7 branes.

In the previous section, we have seen that all models corresponding to the
\-- background are obtained in the T-dual picture from orientifold
groups with the structure $G_{\rm orient.}=Z_N+Z_N\Omega'$. The presence of
the element $\Omega'$ in $G_{\rm orient.}$ induces the appearance of D7
branes.

The analogy between the brane constructions with the \-- and \+-
configurations suggests there must be a natural family of IIB
orientifolds which does not require D7 branes. Indeed,
this is achieved by
constructing a different $\IZ_2$ extension of the orbifold group $\IZ_N$.
If we take an element $\alpha$ generating $\IZ_{2N}$, this can be
constructed as
\beq
G_{\rm orient.}= \{\alpha^{2k}, \alpha^{2k+1}\Omega' \},
\label{grouppm}
\eeq
where $k=0,\ldots,N-1$ \footnote{This orientifold projection has 
appeared in \cite{gj,dp} in the compact case.}. That this is the correct 
structure can be understood as follows. In the absence of NS 
branes, it follows from \cite{witten} that the configuration of two 
oppositely charged O6-planes on a circle is T dual to a type IIB string 
theory compactifed on the dual circle with an orientifold projection 
$\Omega' S$, with $S$ a half shift on the circle \cite{dp}. When the NS 
branes are present, before the orientifold projection the T dual is given 
by a $Z_N$ singularity (more precisely, a $N$-center Taub-NUT space). A 
shift around the whole $U(1)$ orbit in this space is associated to the 
generator $\theta (\equiv \alpha^2)$ of $Z_N$, and a half-shift is 
associated to $\alpha$. In performing the 
orientifold projection, the fact that the O6-planes are oppositely charged 
implies that $\Omega'$ must be accompanied by $\alpha$. The full group we 
have quotiented by is
\beq
G_{\rm orient}= Z_N + Z_N \alpha\Omega'
\eeq
which is equal to (\ref{grouppm}).

In the following we show that D3 brane probes 
on this kind of orientifold singularities actually reproduce the $\NN=2$
theories arising from brane construction in the \+- background.

\subsection{The odd order case}

When the number of NS branes is odd, $N=2P+1$, the orientifold group
(\ref{grouppm}) can also be described  as $G_{\rm orient.}=\IZ_N + \IZ_N R
\Omega' $, where $R$ is a $\IZ_2$ twist inverting the coordinates 6789.

In this case, the absence of D7 branes allows two possible projections on
the D3 branes. We will denote them as the `$Sp$' and `$SO$' projections,
and they correspond to the models (\ref{spec1p}) and (\ref{spec1pp}) of
Section~2,
respectively. Let $\theta=\alpha^2$ generate $\IZ_N$. A choice of D3 brane
Chan-Paton matrices that reproduce the spectra of these theories is
\beqa
&\gamma_{\theta}\, =\, {\rm diag} (1_{v_0},\theta 1_{v_1},\ldots,
\theta^{P} 1_{v_P},\theta^{P+1}1_{v_P}, \ldots, \theta^{2P} 1_{v_1})
\nonumber\\
&{\small \gamma_{R\Omega'} = \pmatrix{
1_{v_0} &  &  &  &  &  & \cr
&  &  &  &  &  & 1_{v_1} \cr
&  &  &  &  & \cdots &  \cr
&  &  &  & 1_{v_P} &  &  \cr
&  &  & 1_{v_P} &  &  &  \cr
&  & \cdots &  &  &  &  \cr
& 1_{v_1} &  &  &  &  &  \cr
} \; ; \;
\gamma_{R\Omega'} =\pmatrix{
\varepsilon_{v_0} &  &  &  &  &  &  \cr
  &  &  &  &  &  & 1_{v_1}  \cr
  &  &  &  &  & \cdots &  \cr
  &  &  &  & 1_{v_P} &  &  \cr
  &  &  & -1_{v_P} &  &  &  \cr
  &  & \cdots &  &  &  &  \cr
  & -1_{v_1} &  &  &  &  &  \cr
}} \nonumber
\eeqa
where $\theta=e^{2i\pi/N}$. The two
possibilities for $\gamma_{R\Omega'}$ correspond to the spectra
(\ref{spec1pp}) and (\ref{spec1p}), respectively.

These matrices are consistent with the group law, and verify the property
\beq
\Tr (\gamma_{\theta^k R\Omega'}^{-1}\gamma_{\theta^k R\Omega'}^T)\; =\;
\pm \Tr (\gamma_{\theta^{2k}})
\eeq
with the upper (lower) sign for the $SO$ ($Sp$) projections, respectively.

\medskip

Since $\Omega'$ is always accompanied by some twist, the final expression
for the tadpoles differs from the familiar ones. The techniques to derive
them, however, are standard \cite{gj,dp} and we just sketch the
computation.

We have the following contributions from the Klein bottle, M\"obius strip
and cylinder
\beqa
\ck & = & \sum_{k=1}^{N-1} \frac{16 \sin^2 [2\pi(k/N+1/2)]}
{[ 4\sin^2 [\pi(k/N+1/2)]]^2} \nonumber\\
\cm & = & \sum_{k=1}^{N-1} (-16) \cos^2 [\pi(k/N+1/2)] \Tr
(\gamma_{\theta^k R\Omega'}^T \gamma_{\theta^k R\Omega'}^{-1}) \nonumber\\
\cc & = & \sum_{k=1}^{N-1} 4 \sin^2 (\pi k/N) (\Tr \gamma_{\theta^k})^2
\eeqa

Notice that we have included the zero mode integration factor in the
denominator of $\ck$, as discussed in \cite{gj}. Also, in the diagrams
involving crosscaps we have included the effects of the $\IZ_2$ twists
$R$, $R_{45}$ in $R\Omega'$. The total contribution is equal to
\beqa
\sum_{k=1}^{N-1} \left[ 4\sin^2 (2\pi k/N) (\Tr \gamma_{\theta^{2k}})^2
\; \mp 16 \sin^2 (\pi k/N) \Tr \gamma_{\theta^{2k}} \;
+\; 4 \frac{\sin^2 (\pi k/N)}{\cos^2 (\pi k/N)} \right]
\eeqa
where the upper (lower) sign is for the $SO$ ($Sp$) projection. It
factorizes as
\beqa
\sum_{k=1}^{N-1} \frac{1}{4\sin^2 (2\pi k/N)} \left[ 4\sin^2 (2\pi k/N)
\Tr \gamma_{\theta^{2k}} \mp 8 \sin^2 (\pi k/N)  \right]^2
\eeqa

The tadpole cancellation conditions are
\beq
4\sin^2 (2\pi k/N) \Tr \gamma_{\theta^{2k}}\mp 8 \sin^2 (\pi k/N) \, =\,
0.
\label{tadpole1p}
\eeq
It is easy to check that, by expanding the sine functions in
exponentials, and recasting the $\Tr \gamma_{\theta^{2k}}$ in terms of the
$v_j$, the tadpole cancellation conditions are equivalent to
\beqa
\sum_{j=0}^{N-1} e^{2\pi i \frac{2k j}{N}} \left[-2v_j
+v_{j-1}+v_{j+1}\pm(4\delta_{j,0}-2\delta_{j,P}-2\delta_{j,P+1})
\right]\,=\,0.
\eeqa
Namely, the finiteness conditions for the $D=4$, $\NN=2$ theory,
eqs. (\ref{beta1pp}), (\ref{beta1p}).

The $Z_3$ example has appeared in \cite{kakush1}. Notice that the indirect
construction technique employed there (considering first a system of D9 
branes and then T dualizing) does not allow the construction of the whole 
infinite family.

The tadpoles for the six-dimensional theory are given by eq.
(\ref{tadpole1p}) after replacing $8\to 32$. The corresponding 
orientifolds yield six-dimensional theories free of anomalies. A similar 
comment applies to the following models.

\subsection{The even order cases}

These families are very interesting from the orientifold point of view,
since the spectra (\ref{spec2p}) and (\ref{spec3p}) suggest there are two
opposite projections acting simultaneously on the D3 branes. We will show 
how the structure of the orientifold group (\ref{grouppm}) allows for 
Chan-Paton matrices yielding these spectra.

\subsubsection{Case with vector structure}

The brane configuration yielding the spectrum in (\ref{spec2p})
is clearly reminiscent of the \-- configuration with spectrum
(\ref{spec2}). This suggests the theory
(\ref{spec2p}) is obtained through an orientifold with possible vector
structure. As discussed in Section~3, this implies certain signs in the
$\IZ_2$ twisted sectors that we should take into account in the tadpole
computation.

The Chan-Paton matrices we use are
\beqa
\gamma_{\alpha^2} & = & {\rm diag} (1_{v_0},\theta 1_{v_1},\ldots,
\theta^{P-1}1_{v_{P-1}},\theta^P
1_{v_P},\theta^{P+1}1_{v_{P-1}},\ldots,\theta^{2P-1} 1_{v_1}) \nonumber \\
\gamma_{\alpha\Omega'} & = & {\small \pmatrix{
1_{v_0}  &  &  &  &  &  &  &  \cr
  &  &  &  &  &  &  & \alpha 1_{v_1} \cr
  &  &  &  &  &  & \cdots &  \cr
  &  &  &  &  & \alpha^{P-1}1_{v_{P-1}} &  &  \cr
  &  &  &  & \alpha^P \varepsilon_{v_P}  &  &  &  \cr
  &  &  & \alpha^{-(P-1)} 1_{v_{P-1}} &  &  &  &  \cr
  &  & \cdots &  &  &  &  &  \cr
  & \alpha^{-1} 1_{v_1} &  &  &  &  &  &  \cr
}}
\eeqa
where $\alpha=e^{2\pi i\over {2N}}$. These matrices verify the group
law, and it is easy to check they give the spectrum (\ref{spec2p}) on the
world-volume of the D3 branes. They also verify the properties
\beqa
\Tr (\gamma_{\alpha^{2k-1}\Omega'}^{T} \gamma_{\alpha^{2k-1}\Omega'}^{-1})
\, = \,  \Tr (\gamma_{(\alpha^2)^{2k-1}}) \, =\, \Tr
(\gamma_{\alpha^{2k+N-1} \Omega'}^{T} \gamma_{\alpha^{2k+N-1}
\Omega'}^{-1})
\label{prop2p}
\eeqa
The positive relative sign between the first and third terms in this
equation agrees with our above comment concerning the vector structure.

\medskip

The contributions from the Klein bottle, M\"obius strip and cylinder, 
are
\beqa
\ck & = & \sum_{k=0}^{N-1} 16 \left[\frac{\cos^2 (\pi \frac{2k-1}{2N})}
{\sin^2 (\pi\frac{2k-1}{2N})} \; + \; 1 \right] \nonumber\\
\cm & = & \sum_{k=0}^{N-1} (-16) \cos^2 (\pi \frac{2k-1}{2N}) \Tr
(\gamma_{\alpha^{2k-1}\Omega}^T \gamma_{\alpha^{2k-1}\Omega}^{-1})
\nonumber\\
\cc & = & \sum_{k=0}^{N-1} 4 \sin^2 (\pi \frac{2k}{2N}) (\Tr
\gamma_{\alpha^{2k}})^2
\label{kmc2p}
\eeqa
We have taken into account that different twists act on diagrams
with and without crosscaps. Notice also the positive relative sign between
the two contributions to the Klein bottle, in correlation with the
positive sign we mentioned concerning (\ref{prop2p}).

Using the property (\ref{prop2p}), the total
contribution can be factorized as
\beqa
\sum_{k=1}^{N-1} \frac{1}{4\sin^2 (\pi k/N)} \left[ 4\sin^2 (\pi k/N)
\Tr \gamma_{(\alpha^2)^k} - 8 \delta_{k,1\; {\rm mod} \; 2}  \right]^2.
\eeqa
The cancellation of the tadpoles reads
\beq
4\sin^2 (\pi k/N) \Tr \gamma_{(\alpha^2)^k} - 8 \delta_{k,1\; {\rm mod}
\; 2}\, =\, 0.
\eeq
These conditions can be seen to be equivalent to the condition of
finiteness for the gauge theory (\ref{beta2p}).

\medskip

\subsubsection{Case `without vector structure'}

The brane configurations yielding the theories in the family
(\ref{spec3p}) are analogous to those realizing the theories
(\ref{spec3}). Thus we expect the T-dual orientifolds to correspond to 
theories without vector structure.

Our choice of Chan-Paton matrices for these cases is
\beqa
\gamma_{\alpha^2} & = & {\rm diag} (e^{\frac{i\pi}{N}}1_{v_1}, 
e^{\frac{i\pi  3}{N}}1_{v_2},\ldots, e^{\frac{i\pi (2P-1)}{N}}1_{v_{P}}, 
e^{\frac{i\pi (2P+1)}{N}}1_{v_P},\ldots, e^{\frac{i\pi(4P-3)}{N}} 1_{v_2},
e^{\frac{i\pi(4P-1)}{N}} 1_{v_1}) \nonumber \\
\gamma_{\alpha\Omega'} & = & {\small \pmatrix{
  &  &  &  &  & e^{\frac{i\pi}{2N}} 1_{v_1}   \cr
  &  &  &  & \cdots &    \cr
  &  &  & e^{\frac{i\pi (2P-1)}{2N}} 1_{v_P}&  &    \cr
  &  & e^{\frac{i\pi (2P+1)}{2N}} 1_{v_P} &  &  &    \cr
  & \cdots &  &  &  &    \cr
e^{\frac{i\pi (4P-1)}{2N}} 1_{v_1}  &  &  &  &  &    \cr
}}
\eeqa
It is a nice exercise to check the spectrum that arises from the
projection is the field theory (\ref{spec3p}).
The matrices satisfy the group law, and have the properties
\beqa
\Tr (\gamma_{\alpha^{2k-1}\Omega'}^{T} \gamma_{\alpha^{2k-1}\Omega'}^{-1})
\, = \, - \Tr (\gamma_{(\alpha^2)^{2k-1}}) \, =\, -\Tr
(\gamma_{\alpha^{2k+N-1} \Omega'}^{T} \gamma_{\alpha^{2k+N-1}
\Omega'}^{-1})
\label{prop3p}
\eeqa
Notice the relative minus sign between the first and third terms.
Consequently, there will be a relative minus sign between the two
contributions to the Klein bottle.

\medskip

The total contribution $\cc + \cm + \ck$ is
\beqa
\sum_{k=1}^{N-1} \left[ 4 \sin^2 (\pi \frac{2k}{2N}) (\Tr
\gamma_{\alpha^{2k}})^2  -16 \cos^2 [\pi \frac{2k-1}{2N}] \Tr
(\gamma_{\alpha^{2k-1}\Omega}^T \gamma_{\alpha^{2k-1}\Omega}^{-1})
+ 16 \left( \frac{\cos^2 [\pi \frac{2k-1}{2N}]}
{\sin^2 [\pi\frac{2k-1}{2N}]} \; - \; 1 \right)\right] \nonumber
\eeqa
Using the properties (\ref{prop3p}), it factorizes as
\beqa
\sum_{k=1}^{N-1} \frac{1}{4\sin^2 (\pi k/N)} \left[ 4\sin^2 (\pi k/N)
\Tr \gamma_{(\alpha^2)^k} + 8 \delta_{k,1\; {\rm mod} \; 2}  \cos (\pi
k/N) \right]^2.
\eeqa

This is very similar to the tadpoles found in \cite{gj,dp}. The only
difference arises from the different twist present in diagrams with and
without crosscaps.

The equations for the cancellation of tadpoles
\beq
4\sin^2 (\pi k/N) \Tr \gamma_{(\alpha^2)^k} + 8 \delta_{k,1\; {\rm mod}
\; 2}  \cos (\pi k/N) \, = \, 0,
\eeq
can be shown to be equivalent to the finiteness conditions (\ref{beta3p}).

\medskip

\section{Final comments}

In this paper we have considered the T-duals of elliptic models with two
O6-planes. They provide the construction of large families of finite
four dimensional $\NN=2$ theories. These orientifolds can also be used
to define anomaly-free six-dimensional field theories with RG fixed
points at the origin of their Coulomb branches.

For the configuration with two negatively charged orientifolds, the
T-duals had already been considered in \cite{bi1}. Our aim has been to
construct the T-duals of the configurations with oppositely charged
orientifold planes. They are realized by type IIB orientifolds which do
not require the presence of D7 branes. The different
families nicely match the different orientifold groups that can be
defined. We have shown that the tadpoles for these orientifolds are
proportional to
the beta functions, so cancellation of tadpoles ensures the finiteness of
the field theories (cancellation of irreducible anomalies in the 
six-dimensional version). 

Notice that the vanishing of the beta functions generically forces the 
different factors in the gauge group to have different rank, so that the 
Chan-Paton matrices are generically not traceless. From the brane picture 
point of view, these ranks can be determined by linking number arguments, 
so the non-tracelessness can be easily tracked down to the O6 and D6 
charges. Thus, a nice result is how 
the tadpole conditions encode the linking numbers of the T-dual brane 
configurations. 

We hope that the realization of these new superconformal field 
theories from D3
branes at orientifold singularities facilitates their study using the recent
developments on the AdS/CFT correspondence, along the lines of
\cite{fth}. 

It would also be very interesting to explore other theories from D3 branes 
at orientifold singularities. We hope that further generalizations, by 
considering non-abelian singularities \cite{bi2}, or considering 
singularities which only preserve 
$\NN=1$ supersymmetry \cite{kakush2, iru}, enlarge the number of these 
extremely interesting field theories.

\bigskip

\bigskip

\bigskip

\centerline{\bf Acknowledgements}

We are grateful to L.~E.~Ib\'a\~nez, K.~Intriligator and A.~Kapustin for 
useful 
discussion. A.M.U. thanks M.~Gonz\'alez for her kind encouragement.
The work of J.P. is supported by the U.S. Department of Energy under Grant 
No. DE-FG02-90-ER40542. The work of A.M.U. is supported by the Ram\'on 
Areces Foundation (Spain).

\end{document}